# Region-Based Rate-Control for H.264/AVC for Low Bit-Rate Applications

Hai-Miao Hu[1], Bo Li[1, 2], Weiyao Lin[3], Wei Li[1, 2], Ming-Ting Sun[4]

1 Beijing Key Laboratory of Digital Media, School of Computer Science and Engineering, Beihang University, Beijing 100191, China

2 State Key Laboratory of Virtual Reality Technology and Systems, Beihang University, Beijing 100191, China

3 Dept. of Electronic Engineering, Shanghai Jiao Tong University, Shanghai, 200240, China

4 Dept. of Electronic Engineering, University of Washington, Seattle, WA 98195

**Abstract** Rate-control plays an important role in video coding. However, in the conventional rate-control algorithms, the number and position of Macroblocks (MBs) inside one basic unit for rate-control is inflexible and predetermined. The different characteristics of the MBs are not fully considered. Also, there is no overall optimization of the coding of basic units. This paper proposes a new region-based rate-control scheme for H.264/AVC to improve the coding efficiency. The inter-frame information is explored to objectively divide one frame into multiple regions based on their rate-distortion behaviors. The MBs with the similar characteristics are classified into the same region, and the entire region instead of a single MB or a group of contiguous MBs is treated as a basic unit for rate-control. A linear rate-quantization stepsize model and a linear distortion-quantization stepsize model are proposed to accurately describe the rate-distortion characteristics for the region-based basic units. Moreover, based on the above linear models, an overall optimization model is proposed to obtain suitable Quantization Parameters (QPs) for the region-based basic units. Experimental results demonstrate that the proposed region-based rate-control approach can achieve both better subjective and objective quality by performing the rate-control adaptively with the content, compared to the conventional rate-control approaches.

# 1 Introduction

With the increasing demand of various video transmission applications, video coding with a high coding efficiency is important. Comparing with exiting video coding standards, H.264 advanced video coding (AVC) [1] can achieve a better coding efficiency under the transmission capacity constraint due to the adoption of several novel coding techniques. In video coding, rate-control plays an important role. The major purpose of rate-control is to determine the quantization stepsizes (or quantization parameters) to regulate the bit-rate produced by the video encoder according to the available channel bandwidth to maximize the video quality. There are many previous works on the rate-control of H.264/AVC [2-23]. Most of them were based on JVT-G012, which was adopted as the rate-control scheme in the H.264/AVC recommended software JM [9, 24].

JVT-G012 introduced a concept of basic unit rate-control. The basic unit is defined to be a group of contiguous MacroBlocks (MBs). The number of MBs in a basic unit is fixed, and it can be a MB, a MB row, a slice, or a frame [9]. The basic unit rate-control is a top-down ordinal process and mainly consists of the following steps. Firstly, the remaining bits for the current frame are allocated to not-yet coded basic units equally. The Mean Absolute Difference (MAD) of the current basic unit is then predicted through the MAD prediction model using the actual MAD of the co-located basic unit in the previous frame. Finally, a quadratic rate-distortion (R-D) model is used to compute the quantization parameter (QP) for the basic unit according to the predicted MAD.

The basic unit rate-control makes the H.264/AVC encoder more flexible and can achieve good performance in video transmission applications, in which the subjective quality of encoded frames is often seriously degraded due to the restricted channel bandwidth. In applications such as videophone and remote video surveillance, the perceptual quality of the foreground moving object is more important than the background. However, given the same QP, the perceptual quality of these regions may be quite different due to the different characteristics of the regions. In order to reduce the transmission bits while keeping good perceptual quality, the features of the human visual system (HVS) are often exploited to allocate more bits to the basic units located in the heuristically determined Regions of Interest (ROI) compared to the other basic units located in the non-ROI [2-5, 25-29].

Although the basic unit rate-control in JVT-G012 gives good performance, its coding efficiency decreases with the decrease of basic unit size. Generally, the MB-layer rate-control results in the worst coding efficiency [6-7]. There are several reasons. Firstly, since the number of MBs in a basic unit inside one frame is inflexible and their locations are predetermined before encoding [7, 9], the correlation between the current basic unit and its co-located basic unit of the previous frame may not be so strong due to local object motions or the global camera motion. Secondly, the basic unit rate-control scheme is a top-down ordinal process without the overall optimization among the basic units in one frame. This may lead to the overdraft of bit-budget for the beginning basic units in a frame and result in bit-starvation for the later basic units. Thirdly, since the MBs inside one basic unit are predetermined regardless of their different characteristics, adopting one uniform model for an entire basic unit cannot produce the optimal result. Although the frame-layer rate-control can achieve better coding efficiency compared to the small size basic unit rate-control, its coding efficiency can be further improved by considering different characteristics of the MBs. The above conclusions will be further confirmed by the experiments in the following sections.

In order to improve the coding efficiency, this paper proposes a new region-based rate-control scheme for H.264/AVC, in which the MBs with similar rate-distortion characteristics are classified into the same basic unit, and an overall optimization is applied to the basic units using a proposed linear rate-quantization stepsize (R-QS) model and a linear distortion-quantization stepsize (D-QS) model. Experimental results demonstrate that the proposed region-based rate-control scheme can achieve both better subjective and objective quality by performing the rate-control adaptively with the content, compared to the conventional basic unit layer rate-control.

Note that in our proposed approach, the numbers of MBs in different basic units are different and the location of MBs belonging to a basic unit may be discontinuous, which is different from conventional basic unit rate-control approaches. Although basic unit rate-control is an active research topic [11, 22-23], to our best knowledge, there has no region-based basic unit rate-control as our proposed approach been proposed before.

The remainder of this paper is organized as follows. Some observations and justifications to support our proposed region-based rate-control are provided in Section 2. Section 3 discusses the proposed region-based basic unit rate-control scheme. The proposed region division algorithm is presented in Section 3.1. The proposed linear R-QS and D-QS models are described in Section 3.2 and the QP determination method is described in Section 3.3. The proposed approach is summarized in Section 3.4. Section 4 presents the experimental results. Some discussions are drawn in Section 5. Finally, the paper is concluded in Section 6.

## 2  Observations and justifications

The proposed region-based rate-control is highly related to three important issues, namely region division, bit-allocation among coding units, and the QP determination. Some observations related to these issues and justifications to our proposed approach are discussed in details as follows.

## A. Region division

In a video frame, a Moving Region (MR) usually attracts more attention. In many previous works [2-5, 25-28], a MR is allocated more bits to keep a good perceptual quality. However, it is difficult to extract reliable MRs. Several techniques [4, 26-27] have been adopted to extract the MRs. These methods are performed after motion estimation (ME), and the QP are adjusted based on the MR extraction result. Since the H.264/AVC reference software uses rate-distortion optimization in the ME which requires the QP information, these MR extraction methods cannot be directly used due to the "region extraction-ME" dilemma. In order to solve this dilemma, difference-based methods [2-3, 5, 25] are proposed which utilize the difference between the current frame and the previous frame. However, when there is a significant temporal variation due to camera motions, these methods cannot produce appropriate results [5, 28].

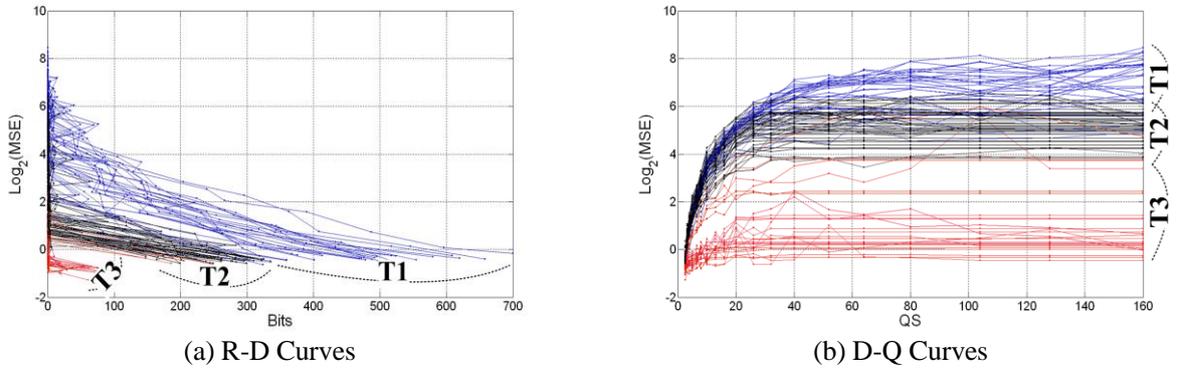

(a) R-D Curves                    (b) D-Q Curves

Fig.1 R-D curves and D-Q Curves of all the MBs in one frame (Grandma, 90[th], QCIF, QP: 12:2:48)

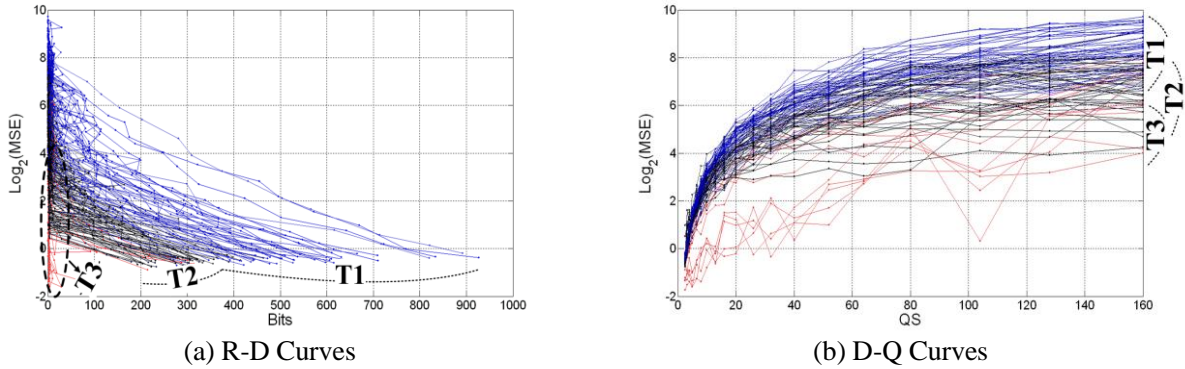

(a) R-D Curves                    (b) D-Q Curves

Fig.2 R-D curves and D-Q curves of all the MBs in one frame (Foreman, 114[th], QCIF, QP: 12:2:48)

We observe that besides attracting more attention, a moving region is more difficult to code and its Rate-Distortion (R-D) characteristic is significantly different from non-moving regions. For a MR, it is not likely that the motion compensation is perfect, and so, there are usually more prediction residuals needing more bits to code given the same QP. The R-D curves of different MBs within a video frame are different, and can be classified into different types with different R-D characteristics. Two examples are shown in Figs.1-2. It can be confirmed that the MBs with the rate-distortion curves labeled as T1 in Figs.1-2 correspond to MRs. These curves have long fat tails, which implies that MRs will consume more bits than the other regions. Also, it can be observed from Fig.1(b) and Fig.2(b) that the coding quality of the MRs is worse than the other regions with the same QP, which may result in a non-uniform quality for the entire frame. The effect is especially significant in the low bit-rate applications, where relative large QPs are utilized.

Moreover, we observe that the characteristics of MBs inside the non-MR may not be uniform as shown in Figs.1-2. The MBs within the non-MR can be classified into two types which are labeled as T2 and T3 in Figs.1-2 with different R-D characteristics. The R-D curves in T3 have shorter tails and descend more rapidly. This implies that the T3 region is easy to code and needs only a small number of bits. The T2 regions are transition regions and have a medium tail. In order to improve the coding efficiency, it is advantageous for the non-MR to be further divided into two sub-regions and to apply different bit-allocation strategies for the different sub-regions according to their characteristics [28, 30].

## B. *Bit allocation among coding units*

Accurate bit-allocation is an important but difficult issue in rate-control in terms of the overall coding performance. In the conventional basic unit rate-control, the remaining bits are equally allocated to the not-yet coded basic units and there is no overall optimization among the basic coding units [9]. However, this top-down ordinal process may lead to overdraft of the bit-budget for the beginning basic units in a frame and bit-starvation for the later basic units as mentioned previously, thus resulting in inappropriate coding results. This is more obvious in low bit-rate applications, where the numbers of bits allocated to each frame are relatively limited. This conclusion is confirmed by the experimental results shown in Table 1, where four video sequences and their corresponding upside-down inverted versions are coded twice with a MB-layer rate-control scheme and a frame-layer rate-control scheme, respectively. The experimental setting is the same as that described in Section 6. We can see that compared to the frame-layer rate-control, the PSNR differences between an original sequence and the corresponding inverted one are much more obvious using the MB-layer rate-control scheme than using the frame-layer rate-control scheme. This implies that the MB-layer rate-control scheme may lead to less stable results even for the same video content. It also shows that the bit-allocation among basic coding units is indeed an issue.

In some rate-control schemes [2-5,25-28], the MBs located in a MR are allocated more bits by adopting larger weighting factors in the bit-allocation strategy than the MBs located in the non-MR. However, the weighting factors are usually constants and determined in heuristic ways which may not achieve good results.

Table 1 Comparison of coding efficiency between the original sequences and the inverted version (i.e., rotate the original frames by 180 degree).

| Sequence (Target bit-rate) | | PSNR (dB) | | Bit-rate (kbps) | |
|---|---|---|---|---|---|
| | | MB-layer[9] | Frame-layer[9] | MB-layer[9] | Frame-layer[9] |
| Claire (16kbps) | Original | 38.17 | 39.35 | 16.05 | 16.04 |
| | Rotated | 38.84 | 39.24 | 16.06 | 16.01 |
| Grandma (16kbps) | Original | 34.97 | 35.87 | 16.05 | 16.00 |
| | Rotated | 35.16 | 35.91 | 16.05 | 15.97 |
| City (24kbps) | Original | 29.04 | 29.56 | 24.08 | 24.03 |
| | Rotated | 29.18 | 29.55 | 24.08 | 24.04 |
| Football (24kbps) | Original | 25.62 | 25.76 | 24.18 | 24.03 |
| | Rotated | 25.33 | 25.81 | 24.11 | 24.06 |
| *Average Absolute Difference* | | *0.32* | *0.05* | | |

## C. QP determination

An R-D model is usually employed to achieve the QP assignment based on the given bit-allocation. Based on different assumptions of the DCT coefficient distribution, many R-D models have been proposed [10-11, 31-33] for rate-control. A logarithmic R-Q model was proposed in [31] by assuming the DCT coefficients have a Gaussian distribution. In [32], an R-D model based on the fraction of zeros among the quantized DCT coefficients was proposed by assuming the DCT coefficients have a Laplacian distribution. Also based on the assumption of Laplacian distribution, a quadratic R-D model was proposed in [33] and was adopted as the rate-control scheme in JVT-G012 for H.264/AVC. Recently, the Cauchy distribution was reported to be a better fit than the Laplacian distribution for H.264/AVC. Several R-D models have been developed for H.264 by using the Cauchy distribution of DCT coefficients [10-12]. However, it is relatively difficult to apply since the mean and variance of Cauchy distribution are not well defined [34]. Moreover, the entropy of the quantized transformed residuals is regarded as the actual rate in the above R-D models. However, the entropy is for the case of independent coding, while the quantized transformed residuals are always dependently entropy coded at the block level, such as the case of run-length coding [34].

The inter-frame correlation is another important issue which affects the accuracy of the R-D model. Since the R-D characteristic of the not-yet coded basic units is not known, the R-D model applied to the current basic unit is established based on the coding results of the co-located basic unit in the previous coded frames. As mentioned, rate-control with a smaller size basic unit cannot achieve as good performance as with a large size basic unit (e.g., the frame-layer rate-control) [6-7]. This is because the correlation between the current basic unit and its co-located basic unit of the previous frame may not be so strong due to local object motions or global camera motions. This statement can be confirmed by the experimental results shown in Fig.3. In Fig.3, the mean square error (MSE) mismatch between different sizes of basic units in the current frame and the co-located basic unit in the previous frame is used as an example to implicitly compare the inter-frame correlation among different layer rate-control schemes. It can be observed that the MB-layer rate-control has the largest estimation error.

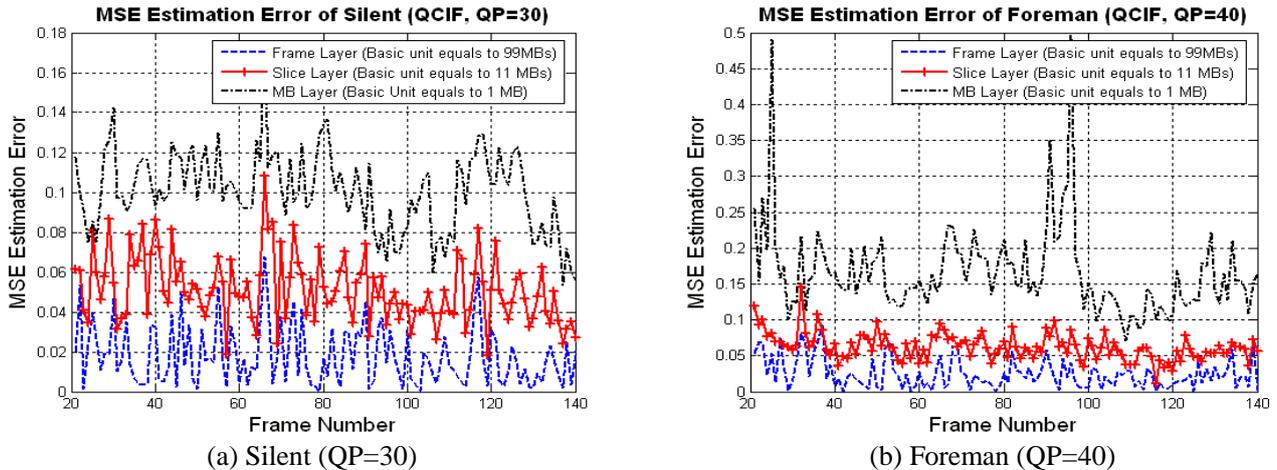

(a) Silent (QP=30)  (b) Foreman (QP=40)
Fig.3 Inter-frame correlation comparison among the different layer rate-control schemes.

## 3 The proposed region-based basic unit rate-control for H.264/AVC

### 3.1 Proposed region division algorithm

The difference-based MR extraction methods are widely utilized and can achieve good MR extraction when the background holds still [3, 5]. However, when there is a significant temporal variation due to the global motions, these methods cannot produce satisfactory results [5]. Therefore, it is reasonable to perform the global motion compensation before the MR extraction.

The Gray Projection Method (GPM) is a simple and effective Global Motion Vector (GMV) estimation method and has been widely used in the digital image stabilization [35]. Its basic idea is to project the two-dimensional image into two independent one-dimensional curves, namely the row projection curve and the column projection curve, and to estimate the GMV by searching the maximum cross-correlation of the projection curves between the current frame and its previous frame. The detailed description of GPM can be found in [35]. Since the GPM utilizes the statistical information, it is robust to local object motions and can achieve good results. Moreover, the GPM has a low computational complexity, which can be easily applied to real-time applications. Therefore, GPM is used in the MR extraction. Based on the estimated GMV, the difference between the current frame and its previous frame for a MB is calculated by Eq.(1):

$$Diff_k(p) = \frac{1}{256} \times \sum_{(i,j) \in p} \left| G_k(i,j) - G_{k-1}(i+GV_k^x, j+GV_k^y) \right| \tag{1}$$

where $Diff_k(p)$ denotes the difference of the $p$th MB in the $k$th frame. $(GV_k^x, GV_k^y)$ represent the horizontal and vertical components of the GMV, respectively. $G_k(i, j)$ denotes the luminance value of the pixel $(i, j)$ in the $k$th frame. Note that $Diff_k(p)$ will be further employed as the complexity measure for the proposed linear R-QS model which will be described in the following section.

The MR is extracted with a predefined threshold based on the calculated difference. Furthermore, according to the HVS feature, the central region of the picture attracts more attention than other parts [2]. Thus, the MBs which have larger temporal differences and are at the central location of the frame are treated as the MR. The proposed MR extraction method is described by Eq.(2), where $MR_k(p)$ equals to 1 if the $p$th MB belongs to the MR, and equals to 0 otherwise.

$$MR_k(p) = \begin{cases} 1, & if \quad \varpi \times \frac{Diff_k(p)}{Diff_k^{avg}} > Th1 \\ 0, & else \end{cases} \quad (2)$$

where $Diff_k^{avg}$ represents the average difference of MBs, $Th1$ is a predefined threshold ($Th1$ is set to 0.75 in this paper), and $\varpi$ is a constant decided by the location of the current MB. In our experiments, $\varpi$ is divided into three levels based on experiments: 1.0 for the central region, 0.1 for the border region, and 0.55 for the transition region. For the QCIF resolution, one MB-wide strip along the edge of the frame is defined as the border region. Two MB-wide strip next to the border region is defined as the transition region. The remaining part is defined as the central region. For the CIF resolution, the strip widths of the border region and transition region are doubled.

The non-MR is sub-divided into the complex region and the flat region based on the observation from Figs.1-2. Due to the "chicken and egg dilemma" described in [9], the MSE information of the current frame is unavailable before the QP value is determined. Since there is no large motion existing in the non-MR, the variation of the content in the non-MR is relative stable. Thus, the MBs in the non-MR have a strong inter-frame correlation and the mean square error (MSE) distribution among the MBs in the non-MR is similar to that in the previous frame. Therefore, the MSE information of the co-located MBs in the previous frames is used as the complex measurement for the non-MR sub-division. A predefined threshold $Th2$ is used to extract the flat region from the non-MR by Eq.(3):

$$NMR_k(p) = \begin{cases} 1, & if \quad \frac{MSE_{k-1}(p)}{MSE_{k-1}^{avg}} > Th2 \\ 0, & else \end{cases} \quad (3)$$

where $MSE_k(p)$ is the actual MSE of the $p$th MB in the $k$th frame, $MSE_k^{avg}$ represents the average MSE and $NMR_k(p)$ is equal to 1 if the $p$th MB belongs to the complex region and is equal to 0 otherwise. $Th2$ is set to 0.5 in this paper.

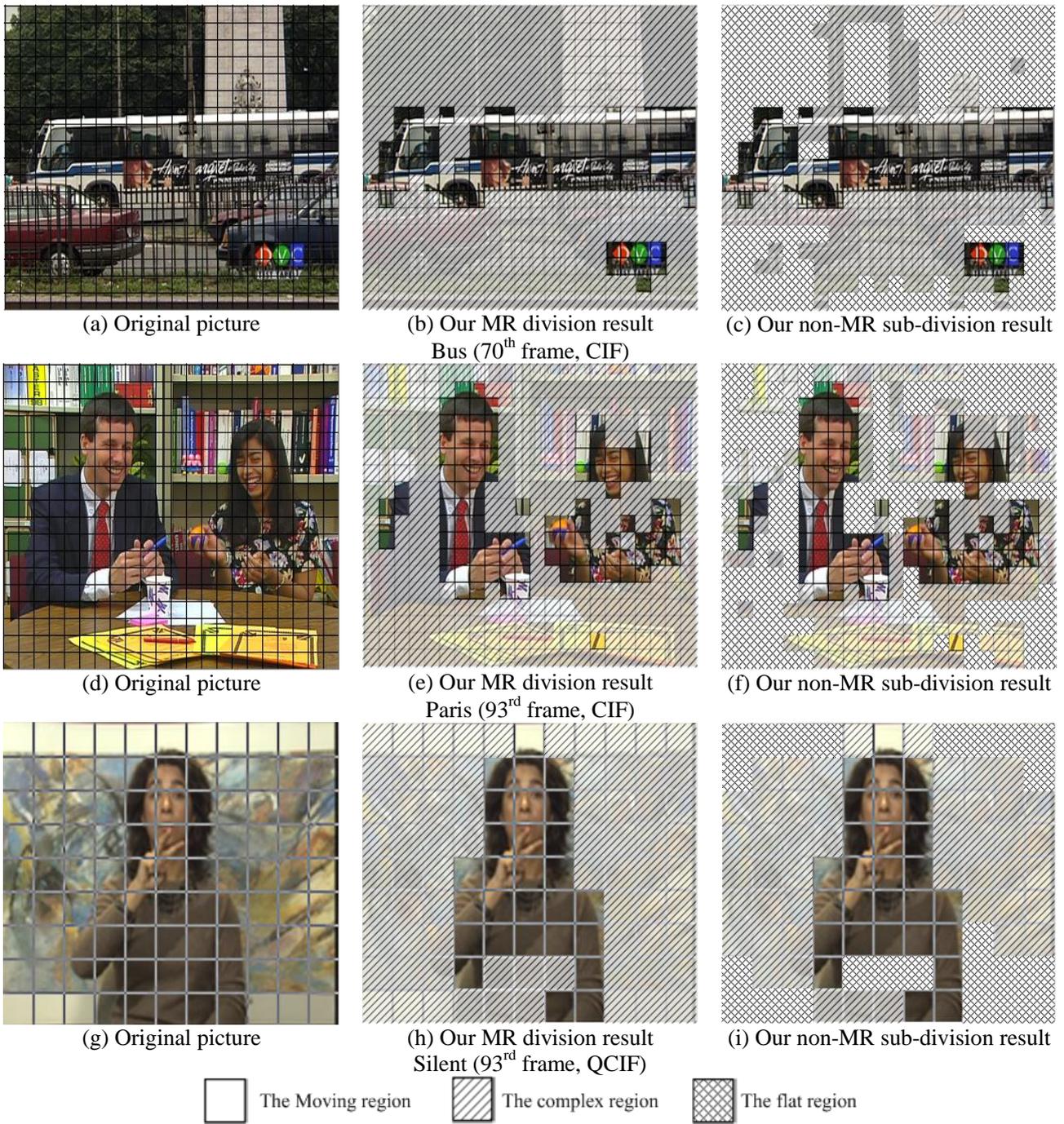

Fig.4 The results of the proposed region division method.

In Fig.4, the "Bus" sequence contains global motions. As shown in the figure, the experimental results demonstrate that based on the GPM, the proposed method can extract the important MR well not only when the background holds still, but also when the background is dynamic. Also, the proposed method can efficiently extract the flat region. Although the extracted MR may not always correspond to the objects completely, it will not affect the coding efficiency. Moreover, with the proposed non-MR sub-division, the remaining part of the object (i.e., object parts that are not included in the MR) usually will be classified to the complex region, but not mixed with the flat region.

In the next section, we will show that the proposed region division method can result in good partitions of MBs with accurate linear R-QS and D-QS models. Note that the predefined thresholds, such as *Th1* and *Th2*, are chosen according to the statistics from the experimental results. Although they can achieve acceptable results for most test sequences, a refining threshold determination method can give better results, as will be discussed in Section 5.

### 3.2 Proposed linear R-QS model and D-QS model

According to the region-division result, we propose to treat the entire region as the basic unit for the rate-control. The proposed scheme includes two important parts, which are described as follows. In our proposed approach, the MBs in each region have a similar R-D behavior, and the MBs in different regions have different R-D behaviors. The R-D models suitable for the region-based basic units should be established correspondingly.

Conventionally, the R-D models are derived based on the statistical properties of video signals and R-D theories [6]. Recently, based on the observation that the distribution of the DCT transform coefficients follows a Cauchy distribution, the R-QS and D-QS relationships are modeled in the form of an exponential function for H.264/AVC [10, 11]. Although Cauchy distribution shows higher accuracy in the case of heavy tails of transformed residuals, it is hard to be applied since the mean and variance of Cauchy distribution are not well defined [34]. Therefore, it is desirable to propose an appropriate R-QS and D-QS model suitable for the region-based basic units.

Several experiments have been carried out to exploit the R-D relationship. The experimental results are shown in Fig.5-6, where "Bits" and "MSE" represent the average bits and the average MSE of MBs in each region, respectively. The R-QS and D-QS models can be represented accurately by linear functions. The accuracy of the curve fitting is specified in Fig.5-6 by the R-squared value (i.e., R2 in the figures), which is an indicator from 0 to 1 that reveals how close the approximated linear function is to the actual data [21]. The R-square values of both the approximated linear R-QS and D-QS models are close to 1, which demonstrates their validity and reliability.

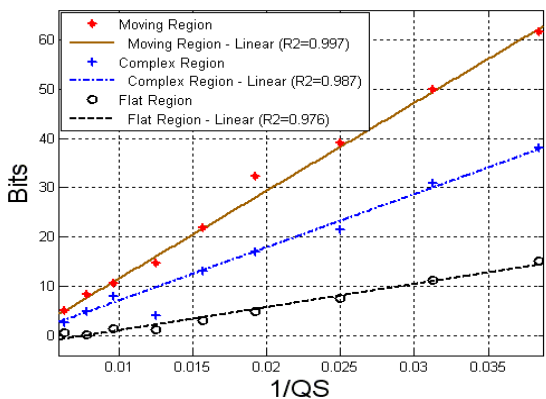
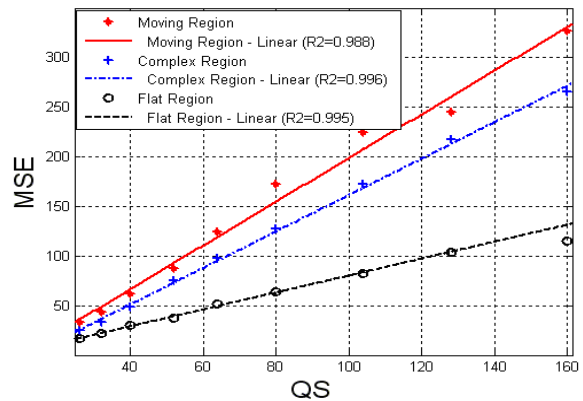

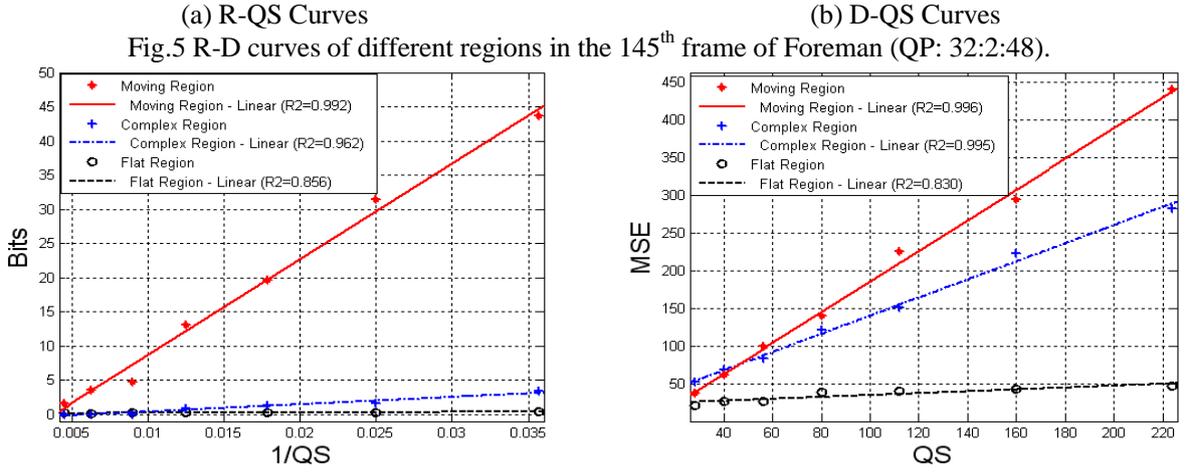

(a) R-QS Curves  (b) D-QS Curves
Fig.5 R-D curves of different regions in the 145[th] frame of Foreman (QP: 32:2:48).

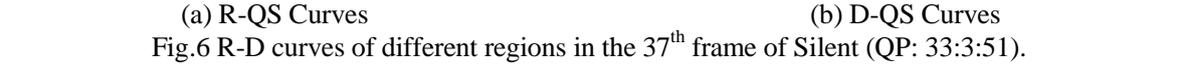

(a) R-QS Curves  (b) D-QS Curves
Fig.6 R-D curves of different regions in the 37[th] frame of Silent (QP: 33:3:51).

In order to make the R-QS model more adaptive to video contents, a parameter related to video complexity, such as the mean absolute difference (MAD), can be associated with the R-QS model [33]. We investigate the relationship between the rate and the mean absolute difference $MAD_{k,r}$ of the *r*th basic unit in the *k*th frame which equals to the average $Diff_k(p)$ of MBs located in the *r*th basic unit (calculated by Eq.(1)). It should be noted that $MAD_{k,r}$ is different from the conventional MAD used in the quadratic R-Q model in JVT-G012 where MAD is predicted based on the actual MAD of the previous frame [9]. Comparing with the conventional MAD, the proposed $MAD_{k,r}$ is also close to having a linear relationship with the bits-rate, which can be seen from Fig.7.

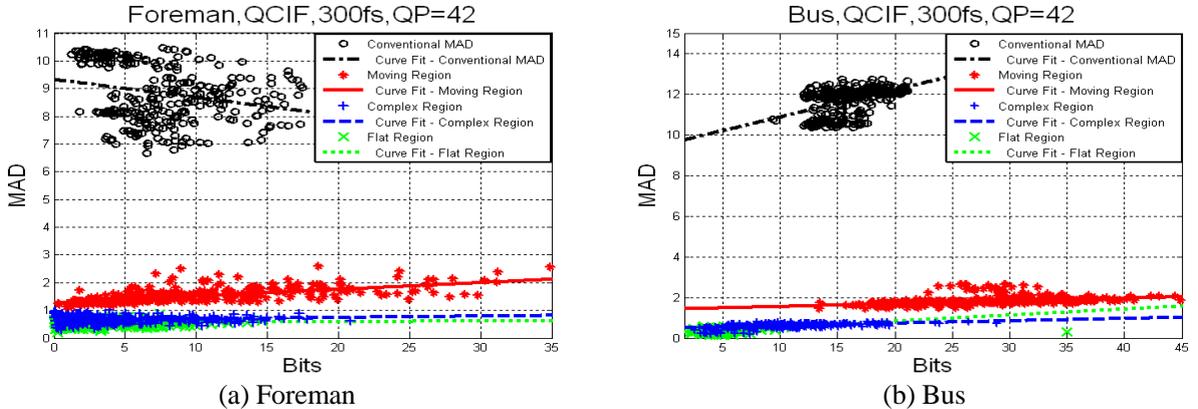

(a) Foreman  (b) Bus
Fig.7 Relationship between bit-rate and MAD (QCIF, 300frames, QP=42) "Bits" represents the average bits of MBs, "Conventional" represents the predicted MAD of one frame with the method in [9].

Therefore, the R-QS and D-QS relationships for each basic unit are modeled in a linear form as follows:

$$\text{R-QS: } R_{k,r} = a_{k,r} \times MAD_{k,r} \times QS_{k,r}^{-1} + b_{k,r}$$
$$\text{D-QS: } D_{k,r} = c_{k,r} \times QS_{k,r} + d_{k,r} \qquad (4)$$

where $R_{k,r}$ denotes the average rate (including both the texture and nontexture information) of MBs located in the *r*th basic unit of the *k*th frame, $D_{k,r}$ is the average distortion of MBs located in the *r*th basic unit of the *k*th frame, and $QS_{k,r}$ is the QS (which will be converted to QP) for the *r*th basic unit of the *k*th frame. $a_{k,r}$, $b_{k,r}$, $c_{k,r}$ and $d_{k,r}$ are parameters of the R-QS and D-QS models and they can be updated through the linear regression method after coding one frame. For the MR, the complex region, and the flat region, *r* equals to *1, 2, and 3*, respectively.

Compared to similar works in [2, 8, 21, 36], our proposed linear R-QS model and D-QS model have the following major differences. Firstly, the proposed models are linear models performed in the quantization stepsize domain. Secondly, the difference between the current frame and the global motion compensated previous frame is directly employed as the complexity measurement of the current frame. This outperforms the conventional methods which use the co-located MB in the previous frame, and can produce a better result when there are camera motions. Thirdly, the impact of the nontexture bits (e.g., headers and overheads such as motion vectors) is significant at low bit-rates. In the proposed R-Q model, the nontexture information is taken into consideration together with the texture information, which can efficiently remove the negative impact from the inaccurate nontexture estimation [21] and can outperform the simple estimation method in [36]. Fourthly, three linear models are established individually for the three basic units, which can efficiently exploit the different R-D characteristics of the MBs. Also, since the MBs with similar characteristics are classified into one basic unit, this method can make full use of the inter-frame correlation to accurately establish the R-D models.

### 3.3 The QP determination

In order to implement an overall bit allocation among the basic units and to minimize the distortion under the given bit allocation, an overall optimization model (as shown in Eq.(5)) is established based on the proposed linear R-QS and D-QS models to obtain the QPs for the three basic units.

$$\arg\min_{QS_{k,r} \in Q} \sum_{r=1}^{3} \left[ N_{k,r} \times \left( c_{k,r} \times QS_{k,r} + d_{k,r} \right) \right]$$
$$s.t. \quad \sum_{r=1}^{3} \left[ N_{k,r} \times \left( a_{k,r} \times MAD_{k,r} \times QS_{k,r}^{-1} + b_{k,r} \right) \right] \leq R_k \quad (5)$$

where $N_{k,r}$ denotes the number of MBs in the *r*th basic unit of the *k*th frame, $QS$ is the set of all possible quantization steps, $R_k$ is the total bit allocation for the entire *k*th frame, which can be obtained through the frame-layer bit allocation method in [9]. The calculated QS will be further converted into QP. The dynamic programming [6, 39] is employed to solve the above problem and to obtain the QPs for the three basic units. The above scheme is labeled as RBUR_T1 (i.e., Region-based Basic Unit Rate-control, Type 1).

Although RBUR_T1 can achieve an appropriate objective quality, the subjective quality, especially for the MR, sometimes does not look good. Since the MR usually attracts more attention and needs more bits, two amendments (namely QS constraint and Lagrange multiplier adjustment) are explicitly added into RBUR_T1 in order to further improve the subjective quality. The modified scheme is labeled as RBUR_T2.

Firstly, for the QS constraint, the optimization model (Eq.(5)) is modified to Eq.(6) to assign a relative small QS (more bits) to the MR, and to assign a relative large QS (fewer bits) to the complex region and the flat region.

$$\underset{QS_{k,r} \in Q}{\arg\min} \sum_{r=1}^{3} \left[ N_{k,r} \times \left( c_{k,r} \times QS_{k,r} + d_{k,r} \right) \right]$$
$$s.t. \quad \sum_{r=1}^{3} \left[ N_{k,r} \times \left( a_{k,r} \times MAD_{k,r} \times QS_{k,r}^{-1} + b_{k,r} \right) \right] \leq R_k \quad and \quad QS_{k,r} \leq QS_{k,r+1} \tag{6}$$

Secondly, the Lagrange multiplier (adopted for mode decision and motion estimation) in H.264/AVC [1, 9] plays an important role for the Rate-Distortion Optimization (RDO). With the cost function $J=D+\lambda \cdot R$ where $D$ represents the distortion, $R$ represents the rate, and $\lambda$ represent the Lagrange multiplier. However, this Lagrange multiplier is only related to QP and no property of the input signal is considered. The Lagrange multiplier should be adjusted to dynamically adapt to different videos as well as different regions [13, 34, 38]. Since the characteristics of different regions are separately represented with the individual linear R-QS and D-QS models, it is reasonable to adjust the Lagrange multiplier corresponding to different regions by taking these models into consideration. On the other hand, the RDO can be treated as a refining process to further keep the perceptual quality of the MR. Since a samller multiplier corresponds to a lower distortion and a higher rate and vice versa [13], a relatively smaller multiplier should be assigned to MR. Therefore, with the consideration of above issues the Lagrange multiplier for mode decision is proposed to be adjusted through the following equations:

$$\lambda_{k,r} = \beta_{k,r} \times \lambda_{k,r}^{Org} = \beta_{k,r} \times \alpha \times 2^{(QP_{k,r}-12)/3} \tag{7}$$

where $\lambda_{k,r}$ is the adjusted Lagrange multiplier for the $r$th basic unit in the $k$th frame, $\lambda_{k,r}^{Org}$ is the original Lagrange multiplier for mode decision in H.264/AVC, $\alpha$ is a constant which is recommended to be 0.85 for H.264/AVC [9], $QP_{k,r}$ is the quantization parameter obtained through Eq.(6). $\beta_{k,r}$ is the calculated by referring to both the adjustment method in [13] and the conventional $\lambda$ determination method (by setting the derivative of the cost function $J$ to zero [34]).

$$\delta_{k,r} = | c_{k,r} / a_{k,r} |$$
$$\beta_{k,r} = N_{k,r} \times \delta_{k,r} / \sum_{i=1}^{3} \left( N_{k,i} \times \delta_{k,i} \right) \tag{8}$$

where $a_{k,r}$ and $c_{k,r}$ are the first-order parameters of the proposed R-QS and D-QS models for the $r$th basic unit in Eq.(4), respectively. $N_{k,r}$ denotes the amount of MBs. The Lagrange multiplier for motion estimation equals to the square root of $\lambda_{k,r}$.

Although the coding efficiency of RBUR_T2 is slightly degraded compared to RBUR_T1, the perceptual quality of RBUR_T2 is improved, which will be confirmed by the experimental results in Section 6.

### 3.4 Summary of the proposed region-based rate-control

The proposed region-based rate-control scheme mainly includes five steps, which are summarized as follows. Note that the other steps of our scheme (e.g., the rate-control for Intra-frames and GOP-layer rate-control) are the same as JVT-G012 described in [9].

*Region Division*

GMP is performed to estimate the GMV. Based on the GMV, the difference between the current frame and the corresponding global motion compensated previous frame for a MB is evaluated by Eq.(1). According to the calculated difference, MBs are classified into three regions, namely the MR, the flat region, and the complex region through Eq.(2) and Eq.(3).

*Frame-layer bit allocation*

The frame-layer bit allocation method of JVT-G012 [9] is employed to determine the target bit for a frame, which is described as follows.

$$R_k = \mu \times \frac{RM_k}{NF_k} + (1-\mu) \times B_k \qquad (9)$$

where $R_k$ denotes the target number of bits of the $k$th frame, $RM_k$ and $NF_k$ are the amount of remaining bits and the number of remaining frames in the current GOP before encoding the $k$th frame, $\mu$ is a constant and is set as 0.5 according to JVT-G012, and $B_k$ is the available bit-rate considering the current buffer occupancy and the target buffer level.

*Optimal QP determination*

In order to implement an overall optimal bit allocation among the basic units and to minimize the distortion under the given bit allocation $R_k$, the optimization model (Eq.(6)) is established based on the proposed linear R-QS and D-QS models described in Eq.(4). In Eq.(6), a QS constraint is explicitly added to differentiate the importance of different regions. The dynamic programming is used to solve Eq.(6) and to obtain the optimal QPs for the three regions. To reduce the blocking artifacts and limit the calculated $QP_{k,r}$ within the range of [1, 51], the final QP of the $r$th basic unit in $k$th frame $QP_{k,r}$ is bounded by:

$$QP_{k,r} = \min\{\max\{QP_{k,r}, QP_{k-1,r} - a, 1\}, QP_{k-1,r} + b, 51\} \qquad (10)$$

where $a$ equals to *3,3,2* and $b$ equals to *2,3,3* for the MR, the complex region, the flat region, respectively.

*Performing RDO*

According to the obtained optimal QP, RDO is performed for each MB in the current frame with the adjusted Lagrange multiplier.

*Parameters update*

After the encoding of each frame, the parameters of the proposed linear R-QS and D-QS models are updated using the linear regression method.

Note that there are several region-based rate-control schemes proposed recently [2-5]. Most of them exploit the features of HVS to allocate more bits to the basic units located in the heuristically determined ROI compared to the other basic units located in the non-ROI. However, the proposed region-based basic unit rate-control is different from the conventional region-based rate-control in the following three aspects.

Firstly, the conventional region-based rate-control schemes are MB-layer rate-control schemes. Since the correlation between the current MB and its co-located MB of the previous frame may not be so strong due to local object motions or the global camera motion, MB-layer rate-control schemes usually result in the worst coding efficiency when compared with the larger size basic unit rate-control schemes. In the proposed method, MBs with similar characteristics are classified into the same region, and the entire region is treated as a basic unit for rate-control. Thus, the proposed method is a kind of "region-based basic unit" rate-control, rather than a "Region of Interest" rate-control.

Secondly, in the conventional region-based methods, the MBs located in a Moving Region (MR) are allocated with more bits by adopting larger weighting factors in the bit-allocation strategy than the MBs located in the non-MR. However, the weighting factors are usually constants and determined in heuristic ways which may not achieve good results. Differently, in order to implement an overall bit-allocation among the region-based basic units, an overall optimization model is established in this paper based on the proposed linear R-QS and D-QS models to obtain the QPs for the three kinds of basic units.

Thirdly, the conventional region-based methods focus on the subjective quality of the ROI, while the overall objective quality may be degraded. However, it is hard to exactly define the ROI. ROI may vary with different applications and different observers. On the other hand, it is hard to find a widely acceptable ROI extraction method. The skin-tone detection method was employed to extract the ROI in [2], which is limited to the conversational video communication applications. The region-based rate-control scheme proposed in [3] (originally proposed for MPEG 4) extracted the ROI only according to the inter-frame difference, which may not work well when there is a camera motion. Differently, the proposed region-based basic unit rate-control scheme is aiming at improving the overall coding efficiency. In the proposed scheme, the region-division criterion is the different characteristics of different MBs, and the MBs with similar rate-distortion characteristics are classified into the same region. The experimental results demonstrate that the proposed region-based rate-control scheme can achieve both better subjective and the overall objective quality (especially, for the MR and the complex region) by performing the rate-control adaptively with the content.

In our previous work [28], we proposed a region-based rate-control scheme using inter-layer information for SVC (Scalable Video Coding). However, it should be noted that this paper is quite different from [28]. Firstly, [28] was mainly aiming at SVC. The main contributions were to fully exploit the inter-layer information to help the moving region (MR) division and the non-MR sub-division. In the work, the rate-control was the MB-layer rate control scheme. However, in this paper, it is not related to exploiting the inter-layer information at all, and a region-based basic unit rate control is proposed to improve the coding efficiency. It is mainly based on the observation that different regions have different RD behaviors and should use different RD models. Secondly, in [28], a group predefined weighting factors were used to allocate different bits to the MBs located in different regions. The R-D model adopted to calculate QP was the conventional quadratic model. However, in this paper, in order to implement an overall bit allocation among the region-based basic units, an overall optimization model is established in this paper based on the proposed linear R-QS and D-QS models to obtain the QPs for the three basic units. So, the contributions of this submitted paper are quite different from those of the previous publication.

## 4  Experiments

In this section, we evaluate the performance of the proposed region-based rate-control scheme for H.264/AVC. The experiments are carried out based on the H.264/AVC reference software JM16.0 [24]. The frame-layer rate-control (which has the largest basic unit, labeled as "FL") and the MB-layer rate-control (which has the smallest basic unit, labeled as "MBL") of JVT-G012 [9] are adopted to compare with the proposed scheme. The test sequences are intra-coded for the first frame (I-frame) and followed with subsequent inter-coded frames (P-frames). The buffer size is set as 0.5*bit-rate. The frame rate is set to 15 fps. RDO and CABAC are enabled. The experimental results are shown in Table 2 and Table 3 and the frame-by-frame PSNR curves are shown in Fig.8, where "RBUR T2" denotes the proposed region-based rate-control scheme and "RBUR T1" denotes the proposed scheme excluding the QS constraint and the Lagrange multiplier adjustment. The average gain in the table is calculated by comparing with the MB-layer rate-control scheme.

Table 2 Comparison of coding performance (QCIF).

| Target bit-rate (Sequence) | | Bit-rate (kbps) | | | | PSNR(dB) | | | |
|---|---|---|---|---|---|---|---|---|---|
| | | MBL | FL | RBUR T1 | RBUR T2 | MBL | FL | RBUR T1 | RBUR T2 |
| 20 kbps | News | 20.14 | 20.02 | 20.07 | 19.99 | 31.63 | 32.09 | 32.73 | 32.47 (+0.86) |
| | Grandma | 20.07 | 20.01 | 19.96 | 19.94 | 35.79 | 36.58 | 36.92 | 36.78 (+0.99) |
| 28 kbps | Container | 28.03 | 28.09 | 28.22 | 28.19 | 36.37 | 36.48 | 37.03 | 36.91 (+0.54) |
| | Salesman | 28.06 | 27.98 | 27.98 | 28.21 | 33.91 | 34.17 | 35.11 | 34.55 (+0.64) |
| 36 | Akiyo | 36.09 | 36.10 | 35.99 | 35.93 | 40.81 | 41.65 | 42.22 | 42.09 (+1.28) |

| kbps | Hall | 36.13 | 36.04 | 36.06 | 36.08 | 37.80 | 37.91 | 38.48 | 38.31 (+0.51) |
| 40 | Mobile | 40.18 | 40.16 | 40.17 | 40.16 | 23.64 | 23.71 | 23.97 | 23.92 (+0.28) |
| kbps | City | 40.11 | 40.06 | 40.12 | 40.10 | 31.03 | 31.21 | 31.37 | 31.31 (+0.28) |
| *Avg. Gain* | | | | | | | *+0.35* | *+0.86* | *+0.67* |

Table 3 Comparison of coding performance (CIF).

| Target bit-rate (Sequence) | | Bit-rate (kbps) | | | | PSNR(dB) | | | |
|---|---|---|---|---|---|---|---|---|---|
| | | MBL | FL | RBUR T1 | RBUR T2 | MBL | FL | RBUR T1 | RBUR T2 |
| 32 kbps | Container | 32.05 | 32.00 | 32.23 | 32.20 | 32.11 | 32.31 | 32.53 | 32.53 (+0.42) |
| | Foreman | 32.20 | 32.62 | 32.89 | 32.68 | 28.73 | 28.87 | 28.98 | 28.89 (+0.16) |
| 64 kbps | Hall | 64.21 | 64.17 | 64.10 | 64.08 | 35.55 | 35.84 | 36.24 | 36.17 (+0.62) |
| | News | 64.32 | 63.95 | 64.31 | 64.03 | 34.50 | 34.94 | 35.42 | 35.19 (+0.69) |
| 96 kbps | City | 96.27 | 95.87 | 96.43 | 96.38 | 29.79 | 30.40 | 30.52 | 30.51 (+0.72) |
| | Akiyo | 96.21 | 96.04 | 96.27 | 96.35 | 41.36 | 42.02 | 42.30 | 42.30 (+0.94) |
| 128 kbps | Paris | 128.51 | 128.07 | 128.03 | 128.21 | 30.39 | 30.46 | 31.51 | 31.30 (+0.91) |
| | Waterfall | 128.27 | 128.28 | 128.17 | 128.45 | 33.80 | 33.85 | 34.34 | 34.25 (+0.45) |
| *Avg. Gain* | | | | | | | *+0.31* | *+0.70* | *+0.61* |

The experimental results demonstrate that our scheme outperforms both the MB-layer and the frame-layer rate-control scheme of JVT-G012. For the QCIF resolution, the proposed scheme (RBUR T2) can gain 0.67 dB and 0.32 dB in the average PSNR when comparing with the MB-layer scheme and the frame-layer scheme, respectively. For the CIF resolution, the proposed scheme can gain 0.61 dB and 0.30 dB in the average PSNR, when comparing with the MB-layer scheme and the frame-layer scheme, respectively. The frame by frame bit allocation and buffer fullness are compared in Fig.9. Note that the bit fluctuation of the proposed scheme is slightly larger than the other two schemes. However, the proposed rate-control can be easily extended to maintain a smooth buffer fullness, which will be discussed in detail in Section 5.

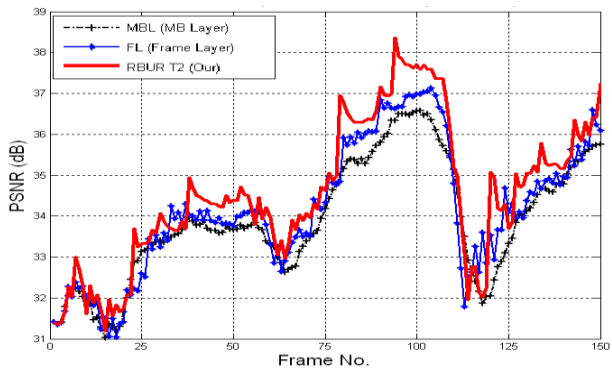 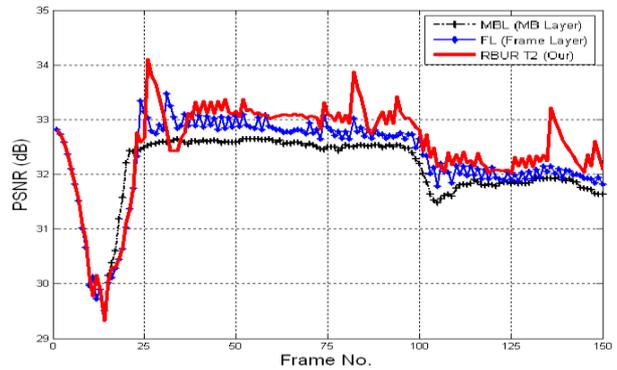

(a) Salesman (QCIF, 28kbps)  (b) Container (CIF, 32kbps)

Fig.8 Frame by frame PSNR curves.

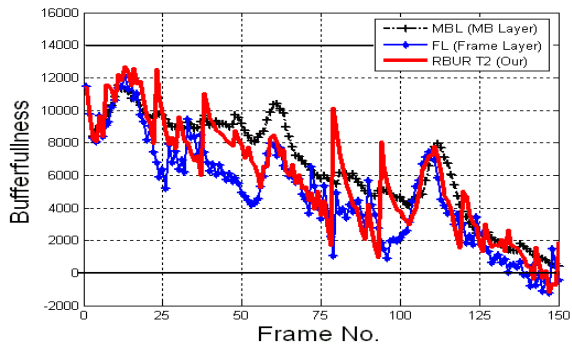 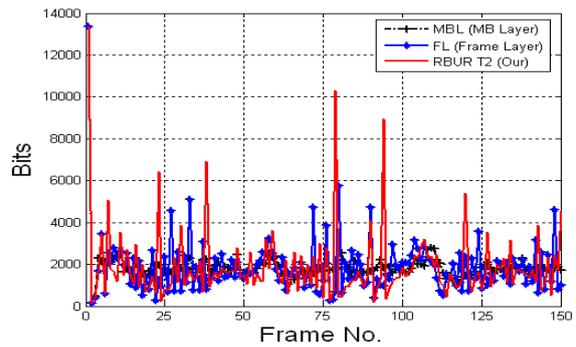

(a) Buffer fullness (Buffer size=0.5*28000)  (b) Bits

Fig.9 Frame by frame bit allocation and buffer fullness (Salesman, QCIF, 28kbps).

Due to adopting the QS constraint and the Lagrange multiplier adjustment, the objective quality (in terms of the PSNR) of "RBUR T2" is not as good as that of the "RBUR T1". It can be observed from Table 2 and Table 3 that comparing with the "RBUR T1", the PSNR gains of "RBUR T2" are degraded by 0.19 dB and 0.09 dB for QCIF resolution and CIF resolution, respectively. However, the subjective quality of "RBUR T2" is better than that of "RBUR T1", which can be observed through some reconstructed frames in Fig.10. When comparing with the frame-layer scheme, both the proposed scheme "RBUR T1" and "RBUR T2" can achieve a better perceptual quality of the man and woman in Fig.10. Moreover, when comparing with "RBUR T1", the perceptual quality of the salesman (especially, the tie of man and the hands and arms of woman) in Fig.10 (d) is further improved.

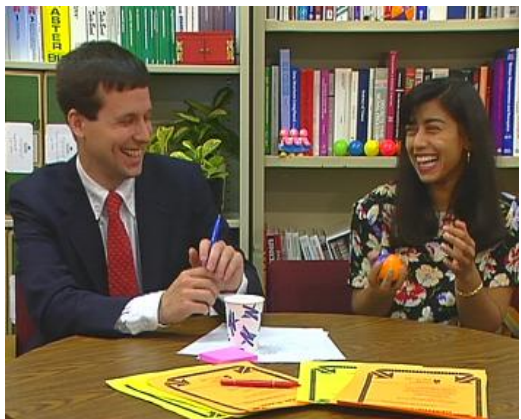

(a) Original

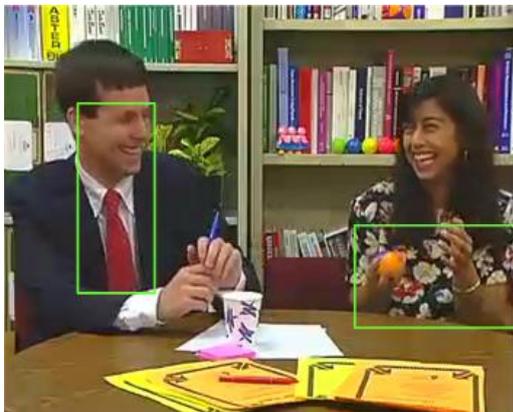 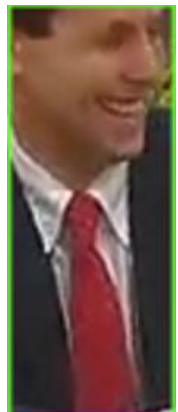 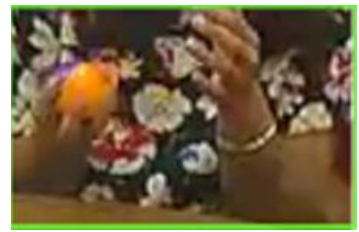

(b) FL (Frame-layer)

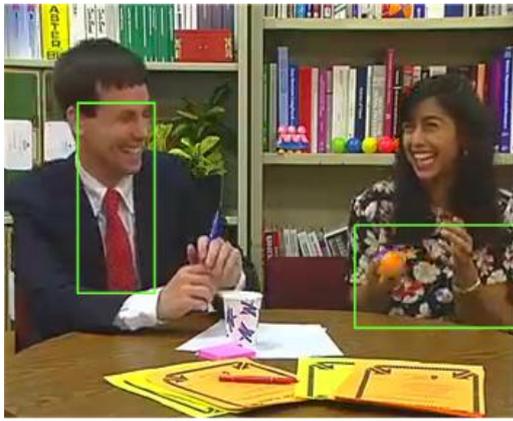 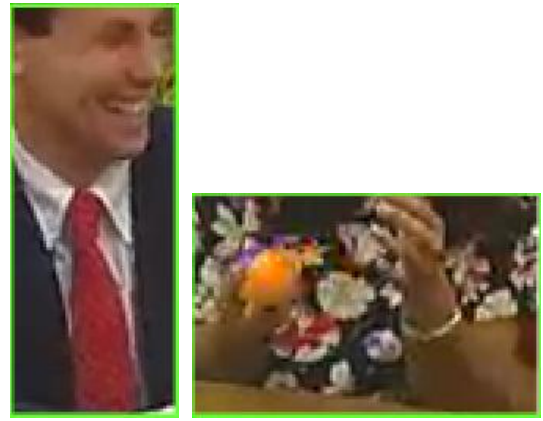

(c) RBUR T1

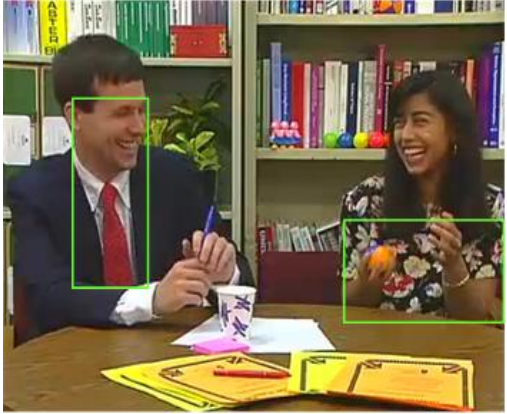 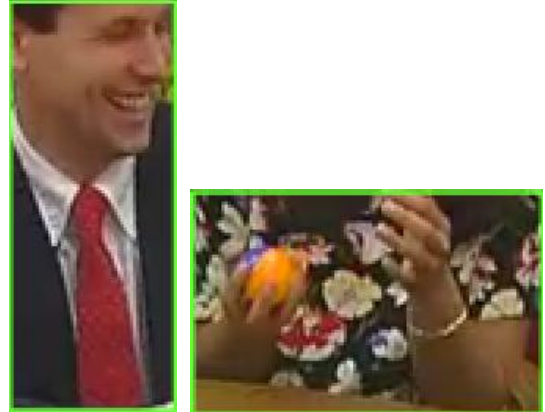

(d) RBUR T2

Fig.10 Comparison of the subjective visual quality of the 38th frame of Paris @ CIF 128kbps.

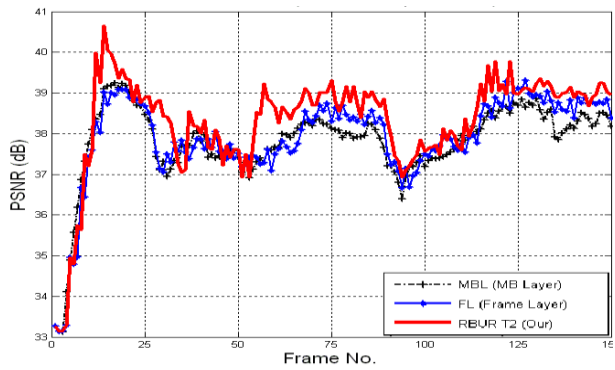 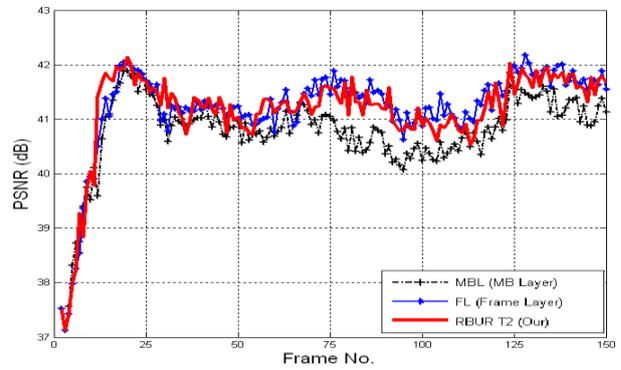

(a) The PSNR curve of the whole picture          (b) The PSNR curve of the Flat Region

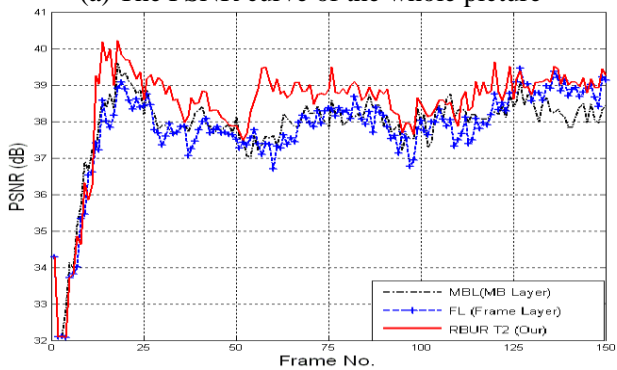 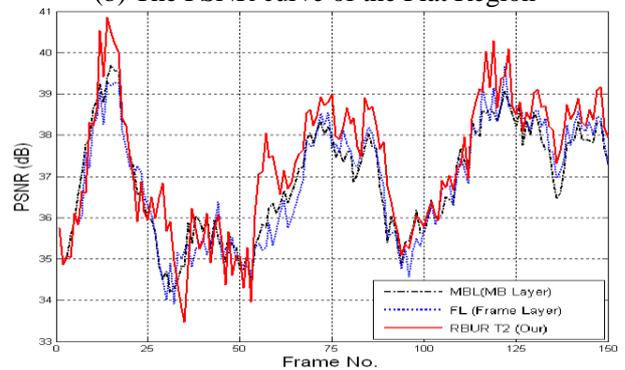

(c) The PSNR curve of the Complex Region      (d) The PSNR curve of the Moving Region

Fig.11 Frame by frame PSNR curves of different regions (Hall QCIF 36kbps, Buffer size = 0.5 * bit rate).

The PSNRs of different regions are compared in Table 4, and the frame-by-frame PSNR curves are shown in Fig.11. The experimental results show that the proposed method can improve the PSNR of both the MR and the complex region while the PSNR gain of the flat region is not as obvious as the other two regions. Compared to the frame-layer rate control scheme, the proposed scheme can gain 0.26 dB and 0.32 dB in the average PSNR for the MR and the complex region, respectively, while the proposed scheme can gain 0.12 dB in the average PSNR for the flat region.

Table 4 PSNR Comparison of different regions (Buffer size = 0.5 * bit rate).

| Sequence | Algorithms | Bit-rate (kbps) | PSNR(dB) | | | |
|---|---|---|---|---|---|---|
| | | | Whole | Flat Region | Complex Region | Moving Region |
| Hall (QCIF, 36kbps) | MBL | 36.13 | 37.80 | 40.49 | 37.87 | 36.89 |
| | FL | 36.04 | 37.91 | 40.93 | 37.79 | 36.87 |
| | RBUR T2 | 36.08 | 38.31 | 40.88 | 38.46 | 37.31 |
| Mobile (QCIF, 36kbps) | MBL | 36.11 | 24.73 | 26.51 | 25.06 | 24.55 |
| | FL | 36.05 | 24.85 | 26.37 | 25.13 | 24.54 |
| | RBUR T2 | 36.23 | 25.04 | 26.50 | 25.37 | 24.74 |
| Container (CIF, 96kbps) | MBL | 96.05 | 34.76 | 39.47 | 33.04 | 35.17 |
| | FL | 96.06 | 35.73 | 40.17 | 35.16 | 35.35 |
| | RBUR T2 | 97.44 | 35.90 | 40.23 | 35.40 | 35.57 |
| Waterfall (CIF, 96kbps) | MBL | 96.33 | 32.49 | 33.76 | 33.16 | 32.23 |
| | FL | 95.89 | 32.68 | 33.09 | 32.91 | 32.49 |
| | RBUR T2 | 96.08 | 33.05 | 33.43 | 33.31 | 32.91 |
| City (4CIF, 512kbps) | MBL | 513.11 | 32.82 | 35.31 | 32.95 | 32.42 |
| | FL | 514.14 | 33.33 | 35.41 | 33.44 | 32.92 |
| | RBUR T2 | 514.58 | 33.50 | 35.52 | 33.64 | 33.12 |
| Mobile (4CIF, 512kbps) | MBL | 514.33 | 25.57 | 30.35 | 25.64 | 24.85 |
| | FL | 511.93 | 25.74 | 30.51 | 25.82 | 24.96 |
| | RBUR T2 | 515.61 | 25.93 | 30.71 | 26.04 | 25.12 |
| Mobcal (1280*720, 1024kbps) | MBL | 1025.55 | 32.47 | 36.07 | 32.56 | 31.73 |
| | FL | 1020.61 | 33.01 | 36.61 | 33.24 | 32.11 |
| | RBUR T2 | 1026.11 | 33.30 | 36.74 | 33.66 | 32.38 |
| Shield (1280*720, 1024kbps) | MBL | 1025.83 | 33.68 | 36.05 | 34.24 | 32.18 |
| | FL | 1022.20 | 34.14 | 36.17 | 34.52 | 32.74 |
| | RBUR T2 | 1018.44 | 34.28 | 36.23 | 34.70 | 32.86 |
| *Avg. Gain* | *FL* | | *+0.38* | *+0.16* | *+0.44* | *+0.24* |
| | *RBUR T2* | | *+0.62* | *+0.28* | *+0.76* | *+0.50* |

Note that the intention of the proposed region-based basic unit rate-control is to fully explore the inter-frame correlation. With the increase of the interval between two frames, the inter-frame correlation is decreased and the improvement of the proposed method is correspondingly decreased. Some experiments are carried out with different Frame Skip Numbers (the number of frames to be skipped in input sequence, e.g., 2 will code every third frame). The result is shown in Table 5. The average gain is calculated by comparing with the frame-layer rate-control scheme. It is obvious that the objective quality of the MR decreases more rapidly than the other two regions. This is because the inter-frame correlation of the MR between the neighboring frames is decreased and the motion compensation for the MR is worse with the increase of Frame Skip Numbers. However, in terms of the overall quality, the proposed method can outperform the frame-layer rate-control regardless of different Frame Skip Numbers.

Table 5 Comparison of coding performance with different Frame Skip Number
(100 Frames, CIF, Buffer size = 0.5 * bit rate).

| Frame Skip Num. | Sequence | Bit-rate (kbps) | | PSNR(dB) | | | | | | | |
|---|---|---|---|---|---|---|---|---|---|---|---|
| | | | | Whole | | Flat Region | | Complex Region | | Moving Region | |
| | | FL | RBUR T2 | FL | RBUR T2 | FL | RBUR T2 | FL | RBUR T2 | FL | RBUR T2 |
| 2 | Hall | 64.15 | 63.99 | 37.84 | 38.00 | 40.20 | 40.38 | 38.20 | 38.56 | 36.37 | 36.35 |
| | Mobile | 128.02 | 128.02 | 27.70 | 27.80 | 30.53 | 30.55 | 27.87 | 28.01 | 27.41 | 27.46 |
| *Avg. Gain* | | | | | *+0.13* | | *+0.10* | | *+0.25* | | *+0.02* |
| 1 | Hall | 63.89 | 64.14 | 36.94 | 37.18 | 39.63 | 39.87 | 37.16 | 37.55 | 35.67 | 35.87 |
| | Mobile | 128.18 | 128.01 | 27.38 | 27.50 | 30.48 | 30.56 | 27.62 | 27.79 | 26.81 | 26.94 |
| *Avg. Gain* | | | | | *+0.18* | | *+0.16* | | *+0.28* | | *+0.16* |
| 0 | Hall | 64.04 | 64.17 | 35.82 | 36.08 | 39.06 | 39.24 | 35.74 | 36.09 | 34.81 | 35.08 |
| | Mobile | 128.39 | 127.77 | 25.80 | 25.96 | 29.54 | 29.64 | 25.94 | 26.13 | 25.29 | 25.48 |
| *Avg. Gain* | | | | | *+0.21* | | *+0.14* | | *+0.27* | | *+0.23* |

Table 6 Comparison of computational complexity (150 Frames, Buffer size = 0.5 * bit rate).

| Resolution (Target bit-rate) | Sequence | Bit-rate (kbps) | | | PSNR(dB) | | | Encoding-Time (ms./f)[*] | | |
|---|---|---|---|---|---|---|---|---|---|---|
| | | MBL | FL | RBUR T2 | MBL | FL | RBUR T2 | MBL | FL | RBUR T2 |
| QCIF 88kbps | Mobile | 88.09 | 88.07 | 88.40 | 28.14 | 28.28 | 28.49 | 391.60 | 391.93 | 397.40 |
| | News | 88.22 | 87.71 | 87.90 | 40.98 | 41.00 | 41.61 | 329.00 | 333.40 | 337.33 |
| CIF 224kbps | Container | 224.10 | 224.84 | 224.05 | 38.42 | 38.60 | 38.91 | 1345.93 | 1349.47 | 1344.87 |
| | Waterfall | 224.39 | 224.24 | 223.81 | 36.09 | 36.10 | 36.49 | 1398.53 | 1401.73 | 1404.27 |
| 4CIF 1536kbps | Piano | 1537.19 | 1536.49 | 1532.05 | 32.08 | 32.31 | 32.37 | 5840.80 | 5830.53 | 5911.73 |
| | Mobile | 1536.06 | 1538.74 | 1535.33 | 29.32 | 29.51 | 29.59 | 6042.93 | 6040.20 | 6123.53 |
| *Avg.* | | | | | | *+0.13* | *+0.41* | | *+0.28%* | *+1.15%* |

When comparing with the conventional basic unit rate-control, the computational complexity of the proposed scheme is increased since the proposed region division process employs GPM to obtain the GMV. More experiments are performed to evaluate the computational complexity and the experimental results are shown in Table 6. Compared to the MB-layer and the frame-layer schemes, the encoding-time of the proposed scheme is increased by 1.15% and 0.87%, respectively. Therefore, the computational complexity increase is negligible. Furthermore, it can be reduced by applying the SKIP Mode early determination for the flat region, which will be discussed in detail in Section 5.

In conclusion, the above experimental results demonstrate that the proposed approaches can achieve both higher subjective and objective quality and can outperform the conventional MB-layer rate-control and the frame-layer rate-control schemes.

---

[*] Average encoding-time by encoding each sequence for 5 times. Hardware platform: CPU: Intel (R) Core (TM) 2 @ 2.66 GHz; Memory: 3.37 GB.

## 5 Discussions

Although the proposed approach is originally aiming at low bit-rate applications, it can be easily extended to applications with a wider range of bit rates. Several experiments have been performed to evaluate the efficiency of the proposed approach in a wider range of bit rates. The buffer size is set to 0.1*bit-rate and 1.0*bit-rate, respectively. The frame-layer rate-control scheme (labeled as "FL") and the MB-layer rate-control scheme (labeled as "MBL") of JVT-G012 [9] are adopted to compare with the proposed scheme (labeled as "RBUR T2"). The experimental results are shown in Tables 7, 8. The average gain is calculated by comparing with the MB-layer rate-control.

Table 7 Comparison of coding performance for a wide range of bit rates (Buffer size = 0.1 * bit rate).

| Sequence (Target bit-rate) | | Bit-rate (kbps) | | | PSNR(dB) | | |
|---|---|---|---|---|---|---|---|
| | | MBL | FL | RBUR T2 | MBL | FL | RBUR T2 |
| Silent (QCIF) | 160kbps | 160.15 | 160.61 | 163.62 | 43.77 | 44.10 | 44.86 (+1.09) |
| | 96kbps | 96.08 | 95.76 | 95.53 | 40.27 | 40.63 | 41.25 (+0.98) |
| | 32kbps | 32.04 | 31.89 | 31.82 | 33.70 | 33.94 | 34.05 (+0.35) |
| Mobile (QCIF) | 160kbps | 160.05 | 160.16 | 160.42 | 30.52 | 30.56 | 30.83 (+0.31) |
| | 96kbps | 96.06 | 96.14 | 95.95 | 28.40 | 28.57 | 28.69 (+0.29) |
| | 32kbps | 32.11 | 32.08 | 32.07 | 24.10 | 24.23 | 24.35 (+0.25) |
| Mobile (CIF) | 384kbps | 384.05 | 383.77 | 382.21 | 29.20 | 29.29 | 29.50 (+0.30) |
| | 256kbps | 256.02 | 255.94 | 257.31 | 28.05 | 28.21 | 28.43 (+0.38) |
| | 128kbps | 128.56 | 128.27 | 128.83 | 25.41 | 25.45 | 25.66 (+0.25) |
| Waterfall (CIF) | 384kbps | 384.09 | 384.25 | 384.26 | 38.35 | 38.47 | 38.69 (+0.34) |
| | 256kbps | 256.21 | 256.58 | 257.40 | 36.59 | 36.71 | 36.99 (+0.40) |
| | 128kbps | 128.47 | 128.08 | 128.48 | 33.73 | 33.95 | 34.21 (+0.48) |
| Mobile (4CIF) | 2048kbps | 2048.03 | 2050.19 | 2048.26 | 30.50 | 30.71 | 30.79 (+0.29) |
| | 1280kbps | 1280.33 | 1282.35 | 1283.55 | 28.63 | 28.80 | 28.89 (+0.26) |
| | 512kbps | 514.27 | 512.68 | 516.24 | 25.43 | 25.65 | 25.81 (+0.38) |
| Mobcal (1280*720) | 3072kbps | 3072.42 | 3072.57 | 3083.73 | 35.18 | 35.41 | 35.49 (+0.31) |
| | 2048kbps | 2049.24 | 2056.48 | 2052.26 | 34.25 | 34.48 | 34.59 (+0.34) |
| | 1024kbps | 1026.52 | 1026.40 | 1026.93 | 32.58 | 32.48 | 33.07 (+0.49) |
| ***Avg. Gain*** | | | | | | ***+0.16*** | ***+0.42*** |

Table 8 Comparison of coding performance for a wide range of bit rates (Buffer size = 1.0 * bit rate).

| Sequence (Target bit-rate) | | Bit-rate (kbps) | | | PSNR(dB) | | |
|---|---|---|---|---|---|---|---|
| | | MBL | FL | RBUR T2 | MBL | FL | RBUR T2 |
| Container (QCIF) | 160kbps | 160.11 | 160.11 | 160.45 | 44.42 | 44.80 | 44.99 (+0.57) |
| | 96kbps | 96.05 | 95.97 | 96.60 | 41.87 | 42.04 | 42.29 (+0.42) |
| | 32kbps | 32.03 | 32.18 | 32.08 | 36.96 | 37.15 | 37.38 (+0.42) |
| News (QCIF) | 160kbps | 160.68 | 159.73 | 160.17 | 45.12 | 45.19 | 45.58 (+0.46) |
| | 96kbps | 96.20 | 95.77 | 95.90 | 41.52 | 41.70 | 42.14 (+0.62) |
| | 32kbps | 32.18 | 32.02 | 32.08 | 34.25 | 34.72 | 34.89 (+0.64) |
| Bus (CIF) | 384kbps | 384.12 | 384.27 | 382.91 | 31.93 | 31.91 | 32.03 (+0.10) |
| | 256kbps | 256.08 | 256.02 | 255.47 | 29.95 | 29.93 | 30.09 (+0.14) |
| | 128kbps | 128.11 | 128.07 | 128.23 | 26.90 | 26.94 | 27.00 (+0.10) |
| Paris (CIF) | 384kbps | 384.39 | 383.94 | 384.03 | 36.98 | 36.90 | 37.48 (+0.50) |
| | 256kbps | 256.55 | 256.10 | 256.30 | 34.06 | 33.89 | 34.98 (+0.92) |
| | 128kbps | 128.40 | 128.22 | 127.99 | 30.42 | 30.49 | 31.35 (+0.93) |
| City (4CIF) | 2048kbps | 2047.60 | 2050.14 | 2049.55 | 37.34 | 37.53 | 37.55 (+0.21) |
| | 1280kbps | 1280.01 | 1283.66 | 1282.54 | 35.86 | 36.13 | 36.18 (+0.32) |

| | 512kbps | 513.11 | 514.14 | 514.58 | 32.82 | 33.33 | 33.50 (+0.68) |
|---|---|---|---|---|---|---|---|
| Stockholm (1280*720) | 3072kbps | 3072.12 | 3077.63 | 3074.56 | 35.75 | 35.85 | 35.82 (+0.08) |
| | 2048kbps | 2048.19 | 2052.01 | 2049.99 | 35.18 | 35.30 | 35.31 (+0.13) |
| | 1024kbps | 1024.57 | 1024.02 | 1025.55 | 34.24 | 34.50 | 34.51 (+0.27) |
| *Avg. Gain* | | | | | | *+0.15* | *+0.42* |

The experimental results demonstrate that the proposed scheme can also obtain a better coding result in a wider range of bit rates. Comparing with the MB-layer rate-control scheme, the proposed approach can gains 0.42 dB in average PSNR under the different delay constraints, respectively. When compared to the frame-layer rate-control scheme, the proposed approach can gain 0.26 dB and 0.27 dB in average PSNR under the different delay constraints, respectively.

Although the computational complexity of the proposed rate-control scheme is increased compared to the conventional basic unit rate-control, the computation complexity increase is negligible and it can be reduced with the consideration of following two issues. Firstly, in terms of the mode decision, the SKIP mode occupies a large proportion, especially in low bit-rate applications. Several statistical information is listed in Tables 9, 10. It can be observed that around 90% of the MBs located in the flat region adopt the SKIP mode. Therefore, over a half SKIP mode distributes in the flat region which is a less important region as the other two regions. Therefore, a SKIP mode early determination process can be implemented for the flat region, which can reduce the computational complexity. Secondly, the obtained GMV can be referred by the motion estimation process [37]. This can further reduce the computational complexity.

Table 9 SKIP mode distribution ratio for three regions (QCIF, Constant QP=40).

| Sequence | SKIP mode distribution ratio (%) | | | Prediction Accuracy of SKIP mode for Flat Region (%) | Coding Results | |
|---|---|---|---|---|---|---|
| | MR | Complex Region | Flat Region | | Bit rate (kbps) | PSNR (dB) |
| Silent | 0.17 | 0.42 | 0.41 | 0.99 | 9.53 | 29.27 |
| News | 0.16 | 0.30 | 0.54 | 0.99 | 10.64 | 29.57 |
| Miss-America | 0.32 | 0.08 | 0.59 | 0.99 | 4.4 | 33.47 |
| Mother-daughter | 0.29 | 0.23 | 0.48 | 0.94 | 4.95 | 30.82 |
| Carphone | 0.29 | 0.22 | 0.48 | 0.94 | 9.72 | 29.44 |
| Table | 0.11 | 0.26 | 0.63 | 0.86 | 20.62 | 28.54 |
| Football | 0.09 | 0.24 | 0.67 | 0.72 | 59.04 | 28.43 |
| *Avg.* | *0.21* | *0.25* | *0.54* | *0.93* | | |

Table 10 SKIP mode distribution ratio for three regions (QCIF, Constant bit-rate=32kbps).

| Sequence | SKIP mode distribution ratio (%) | | | Prediction Accuracy of SKIP mode for Flat Region (%) | Coding Results | |
|---|---|---|---|---|---|---|
| | MR | Complex Region | Flat Region | | Bit rate (kbps) | PSNR (dB) |
| Silent | 0.08 | 0.55 | 0.37 | 0.95 | 32.09 | 33.94 |
| News | 0.11 | 0.46 | 0.43 | 0.98 | 32.07 | 34.53 |
| Miss-America | 0.10 | 0.14 | 0.64 | 0.87 | 31.94 | 42.20 |
| Mother-daughter | 0.15 | 0.28 | 0.57 | 0.95 | 32.08 | 37.76 |
| Carphone | 0.19 | 0.24 | 0.58 | 0.81 | 31.94 | 34.63 |
| Table | 0.09 | 0.34 | 0.57 | 0.78 | 32.01 | 30.42 |

| Football | 0.17 | 0.24 | 0.59 | 0.77 | 32.06 | 26.58 |
|---|---|---|---|---|---|---|
| *Avg.* | *0.13* | *0.32* | *0.53* | *0.87* | | |

It can be observed from Fig.12 that the bit-rate of the MR has a larger fluctuation than the other two kinds of regions. This implies that the bit-rate fluctuation of the whole picture mainly comes from the MR. In compliance with the proposed region-based basic unit, a low-delay rate-control can be implemented by separately carrying out the different processes for different regions, such as the MAD prediction process, which is important for the low-delay rate-control. As for the MR, the switched MAD prediction method proposed in [21] can be employed to enhance the traditional linear MAD prediction model [9], which is not suitable for predicting abrupt MAD fluctuations. As for the other two regions, the linear MAD prediction may achieve an accurate prediction results since there is no high motion existing in these two regions. Therefore, the proposed region-based basic unit rate-control can be slightly modified to justify the low-delay constraints.

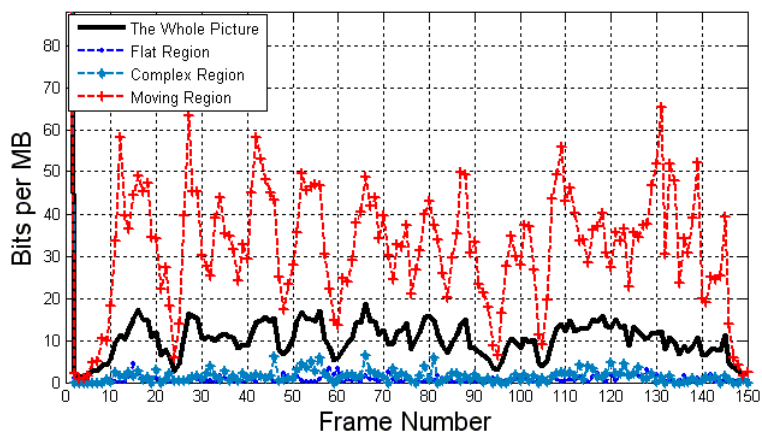

Fig.12 The average bits of different regions (Silent QCIF QP=36).

Although the thresholds used in the region division can achieve acceptable results for most test sequences, a refining threshold determination method is desired. These thresholds could be updated according to different contents and a more complicate threshold determination method can be implemented (e.g., taking more HVS features into the consideration or employing video content analysis methods) to achieve a more accurate region division. Moreover, some mathematical methods, such as Cluster Analysis, can also be included to achieve more accurate MB classification. Obviously, the above methods may result in computational complexity increase. However, they may further improve the coding efficiency and could be one of our future works.

## 6  Conclusion

In this paper, we propose a region-based rate-control scheme for H.264/AVC. The video frames are divided into three regions according to their different characteristics. Each entire region is treated as a basic unit, and a linear R-QS model and a linear D-QS model are proposed for different basic units. Based on this, an optimization model is established to calculate the optimal QPs for the three basic units. Two amendments are added to achieve not only a good objective quality but also a high subjective quality. Experimental results demonstrate that the proposed region-based rate-control outperforms the conventional basic unit rate-control schemes. Comparing with the MB-layer scheme and the frame-layer scheme, the proposed scheme can gain around 0.6 dB and 0.25 dB in the average PSNR, respectively.

Although the proposed scheme is originally aiming at the low bit-rate applications, it can be easily extended to applications with a wider range of bit-rates. Moreover, the region-based scheme can also be easily extended to smooth the bit-rate fluctuation, which is important for low-delay applications. More complicated thresholds determination methods could also be employed to achieve a more accurate region division, which may improve the coding efficiency further.